\documentclass[11pt,twoside]{article}

\usepackage{ihep,epsfig,amsfonts,amsmath,amssymb,amsbsy}

\usepackage{graphicx}
\usepackage[cp866]{inputenc}
\usepackage[russian]{babel}

\usepackage{graphicx} 

\usepackage[russian]{babel}
\usepackage{caption2}

\usepackage {graphics,longtable}

\begin{document}

\begin{center}
{\Large\bf SEARCHES FOR BLACK HOLES. THE MOST RECENT RESULTS}
\end{center}

\begin{center}
{\large\bf A.M.Cherepashchuk}
\end{center}

\begin{center}
{\large\it Sternberg Astronomical Institute, Moscow, Russia}
\end{center}

\bigskip

\begin{center}
Abstract
\end{center}

{\small Black holes were predicted by Einstein General Relativity (GR).
Because of unusual properties of these objects their existence is almost unbelievable. There are gravitation
theories which do not predict the
black hole appearance. By now, astronomers discovered a few hundreds of massive and highly compact objects
with observational features which are very similar to the properties of black holes predicted by GR. To confirm
existence of black holes in the Universe, a number of groundbased and space experiments are planned
to be held in the coming decade.}

\section{Introduction}

The story of searching for black holes is not completed yet. Although it is worthwhile emphasizing that
about 200 massive and  highly compact objects have been discovered so far which possess properties very
similar to those of black holes. A black hole (BH) is an object which has an escape velocity equal to
the speed of light in vacuum, $c=300,000$ km/s. Conception of black hole arose after discovery of Newton's law
of universal gravitation in 1687. In 1783 John Michell expressed an idea of dark stars having
gra\-vi\-ta\-tio\-nal
field so strong that even the light cannot escape outwards. The same idea was put forward in 1798 by
Pierre-Simon Laplace. The existence of black holes is predicted in Einstein's General Theory of Relativity
(GTR). A characteristic size of a black hole is defined by the Schwarzschild radius (gravitational radius),
$r_g=2GM/c^2$, where M is a mass of the object, c is the speed of light, $G=6.67\cdot 10^{-8} \mbox{dyn}
\cdot \mbox{cm}^2 \mbox{g}^{-2}$ is the gravitational constant.

The gravitational radius is equal to

$$
r_g=\left\{
\begin{array}{lllll}
9 & \mbox{mm} & \mbox{for} & \mbox{the Earth} & (M=6\cdot 10^{27} \,\mbox{g})\\
30 & \mbox{km} & \mbox{for} & 10 M_{\odot} & (M=2\cdot 10^{34}\, \mbox{g})\\
40 & \mbox{AU} & \mbox{for} & \mbox{$M=2\cdot 10^9 M_{\odot}$} & (M=4\cdot 10^{42}\, \mbox{g}),\\
\end{array}
\right.
$$

where $\odot$ is a symbol for the Sun, 1 Astronomical Unit (AU) is the Earth's average distance from
the Sun, $1.5\cdot 10^{13}$ cm.

The boundary of a black hole is identified with the event horizon where time is stopped for a distant
observer. Therefore, all the events occuring under the event horizon are beyond the reach of the
observer. The radius of the event horizon is equal to the gravitational radius for a non-rotating
(Schwarzschild) black hole and is less than the gravitational radius for a rotating black hole. In this case
the event horizon is located inside the black hole's ergosphere where the vortical gravitational field is
created. It is John Wheeler who called these objects "black holes" (BH) in 1968.

Properties of BH are described, for example, in ~\cite{NF86} as well as in a recent review
by the same authors~\cite{NF01}.It is worth mentioning that the event horizon is not a solid observable
surface. It can be removed with the use of a proper reference frame. For example, there is no event horizon
at all
for an observer freely falling onto a BH. Such an observer could get into the BH and see
the central singularity which contains all the compressed matter but he could transmit no information outward.
Because of extraordinary properties of BH their existence in the Universe is doubted and has been
debated for some decades. Some versions of the gravitational theory deny BH (see, e.g.,~\cite{Log01}),
the problem of searching
for them getting more intriguing and interesting. Moreover, because of relativistic effect of time
retardation near the event horizon, "contemporary" BH do not seem to have the event horizons
completely formed. Astronomers consider these objects to be "virtual" BH with "virtual"
event horizons.

In 1964 a pioneering work by Ya.B.Zel'dovich~\cite{Z64} and E.E.Salpeter~\cite{S64} appeared where
the possibility to observe black holes was predicted due to the powerful energy release in the process
of non-spherical accretion of matter to the BH. In~\cite{ShS73, PR72, NT73} a theory of disk accretion
to neutron stars (NS) and BH
was developed which was very helpful in understanding the nature of compact X-ray sources discovered aboard
UHURU satellite~\cite{FJC78} and identifying them as accreting NS and BH in stellar binary systems. Optical
in\-ves\-ti\-ga\-tions of X-ray binary systems~\cite{ChEK72, BB72, LSCh73} stimulated development of reliable
methods for determining NS and BH masses~\cite{GRCh91}. 3D gasdynamic models of gas flow in binary systems
made mechanisms of accretion disk formation more clear~\cite{BBK97}. Models of advection-dominated disks
around BH proposed
in~\cite{R82, NMcCY96} account for anomalously low luminosity of accreting BH in nuclei of
normal galaxies and low-mass X-ray binary systems. Searching for stellar-mass BH being rather succesful,
a certain progress is likely to have appeared recently in researching super-mass BH in galactic nuclei.
The most convincing evidence for super-mass compact objects has been obtained recently while investigating
"quiescent" ga\-lac\-tic nuclei (see, for example, the recent Symposium
results~\cite{KvdHW01}
and the
review~\cite{Ch03}), although  quasars and
nuclei of active galaxies were considered to be the first super-mass BH candidates
~\cite{S64, ZN64, L69}.

\section{Methods for Searching Black Holes}

Three types of BH are suggested to exist:

\begin{enumerate}
\item Stellar mass BH, $M=(3-50)M_{\odot}$, which are formed at advanced stages of massive star evolution.
If an evolved stellar core has got a mass $M_c \le(1.2-1.4)M_{\odot}$ it yields a white dwarf, if its final
mass  is $M_c<3M_{\odot}$ it produces a neutron star, and if the mass is $M_c \ge 3M_{\odot}$ a BH is likely
to be the final object.

\item Super-mass BH in galactic nuclei, $M=10^6 - 10^9 M_{\odot}$.

\item Primary BH formed at early stages of the evolution of the Universe.
\end{enumerate}

As to intermediate-mass BH with masses $M=(10^2 - 10^4)M_{\odot}$, their existence is still debated. Whether or
not they can appear is not clear yet. Actually a number of stellar mass BH has been found to be nearly
20, and a number of super-mass BH in galactic nuclei around 200.

Two main problems are to be solved in the process of searching for BH:

\begin{enumerate}
\item Looking for massive, $M>3M_{\odot}$, compact objects as possible BH candidates.

\item Looking for sufficient observational criteria for them being real BH.
\end{enumerate}

Astronomical observations of BH are possible owing to X-ray halos around them occuring in non-spherical
accretion of
matter~\cite{Z64, S64, ShS73, PR72, NT73}.
But terrestrial atmosphere is opaque for
X-rays. It is in 1962 that the first X-ray source beyond the solar system, Sco X-1, was discovered aboard
an American Aerobee rocket in an experiment leaded by Riccardo Giacconi who was awarded the Nobel Prize in 2002.

Astronomical observations of BH involve three steps.
\begin{enumerate}
\item Analysing  motions of "trial bodies" (stars, gas clouds, gas disks) in the BH gra\-vi\-ta\-tio\-nal fields
provides estimates of BH masses. Characteristic distances being much greater than gravitational radii,
Newton law may be used to make such estimates.

\item BH radii are measured by indirect approaches: through analyses of fast X-ray variability, X-ray line
profiles, etc.

\item The most challenging problem is to look for observational evidence of the event horizon.
\end{enumerate}

\begin{figure}[th] 
\centering\includegraphics[width=9cm]{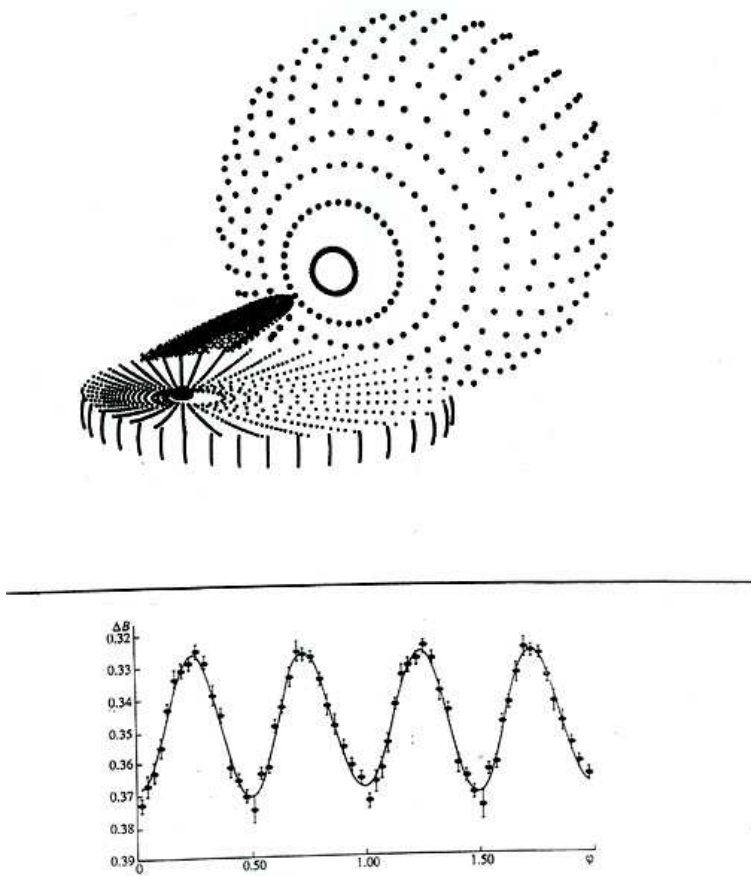}
\caption{Accretion of matter onto NS or BH in the inner parts of
the accretion disk. A model of
an X-ray binary system is shown at the top (\textbf{sight from above}).
Accretion onto a fast rotating NS with strong magnetic field
results in X-ray pulsar phenomenon (shown in the middle of
\textbf{Fig.1}).
 The X-ray pulse period ($p=1.24$ s) and the
pulsation phases of the Her X-1 pulsar remain stable over the
period of 30 years. This indicates that the NS has a solid observable surface.
In the case of accretion onto a BH (at the bottom of the
\textbf{Fig.1}),
 X-ray pulses are not observed since there is no
any observable surface. Only irregular variability of X-ray
radiation is present at time scales up to $t_{min} \sim r_g/c \sim
10^{-3}\div 10^{-4}$ s.}
 \label{Fig_1}
\end{figure}

A final proof whether or not an object is a BH can be obtained if
a reliable event horizon or an ergosphere is revealed (for a
rotating BH). Special groundbased and space projects are planned to
solve this significant problem. The main observational criteria of
an accreting stellar mass BH are great mass, powerful X-ray
radiation, and absence of X-ray pulses or 1st type X-ray bursts.
Pulses or bursts are peculiar to accreting NS which have got
observable surface and fast rotation. In a strong magnetic field
($~10^{12}$ Gs) matter from the inner parts of the NS accretion
disk is tunnelled via magnetic lines of force to the NS magnetic
poles and, after having been collided with the NS surface, is
heated up to temperatures over 10 million degrees, up to some
hundred million degrees. Hot spots are formed in the areas of
collision. Since the rotational axis is not coincident with the
magnetic dipole axis, these hot X-ray spots on the NS surface are
alternately visible to an Earth observer or shielded by the NS
body (``lighthouse effect''). Thus a phenomenon of X-ray pulsar
appears when a strictly periodic X-ray radiation comes out,
periods lying between less than a second and some minutes
(Fig.~\ref{Fig_1}).

\renewcommand{\figurename}{Fig.}

If the NS magnetic field is rather small (less than $10^8$ Gs)
matter from the inner parts of the accretion disk is spreaded
about and accumulated on the NS surface resulting in nuclear
explosion of accumulated matter. Thus a phenomenon of the 1st type
X-ray burster appears, i.e. short (for some seconds) and powerful
X-ray bursts. So, X-ray pulses and the 1st type X-ray bursts are
considered to justify that an accreting relativistic object has
got the observable surface. Another evidence of such observable
surface is when a relativistic object displays pulses in radio
occuring due to ejection of charged relativistic particles in a
strong magnetic field of a fast rotating NS. In this case, short
(between milliseconds and seconds) and strictly periodic pulses of
radio emission come out from the NS.

X-ray pulses, radio pulses or the 1st type X-ray bursts are not likely to occur on an accreting BH.
What may be expected, however, is just X-ray irregular variability on time scales
$~r_g /c \cong 10^3 - 10^4$ Fig.~\ref{Fig_2}.

Sometimes NS may be observed which display neither X-ray pulses, radio pulses, nor the 1st type X-ray bursts.
Hence, their absence is a necessary but insufficient condition to identify a compact object with a real BH.

There are no so far sufficient observational criteria to select a BH. It is worthwhile mentioning, however,
that all the necessary criteria taking into account GR-effects are fullfilled for all known BH candidates
(~200). So modern astronomers, with a kind of forced argumentation, call the BH candidates "black holes".

\section{Stellar mass Black Holes}

There are two types of X-ray binary systems which can contain BH (see the Catalog~\cite{ChKKh96}):

\begin{enumerate}

\item Quasistationary X-ray binaries with massive hot companion stars.
\item Transient (flashing) X-ray binaries, i.e. X-ray novae, with low-mass cool companion stars.

\end{enumerate}

About a thousand X-ray binary systems in the Milky Way and near galaxies have been discovered so far
owing to specialized X-ray observatories launched into orbits around the Earth. Noticeable contribution
into discovery and researches of X-ray binary systems was made by the Soviet and Russian Mir Kvant and Granat
X-ray observatories under R.A.Sunyaev (see, for example,~\cite{SLG88, SChG91}).
Successful observations of BH candidates in a hard X-ray spectrum that is the most appropriate range to search
for
BHs~\cite{ChSS03} have been carried out by the international X-ray and gamma observatory INTEGRAl
launched by a Russian Proton
carrier rocket in October 2002~\cite{ChSS03}. The Russian scientific co-leader of the project is R.A.Sunyaev.

\begin{figure}[p] 
\centering\includegraphics[width=16cm,scale=0.7] {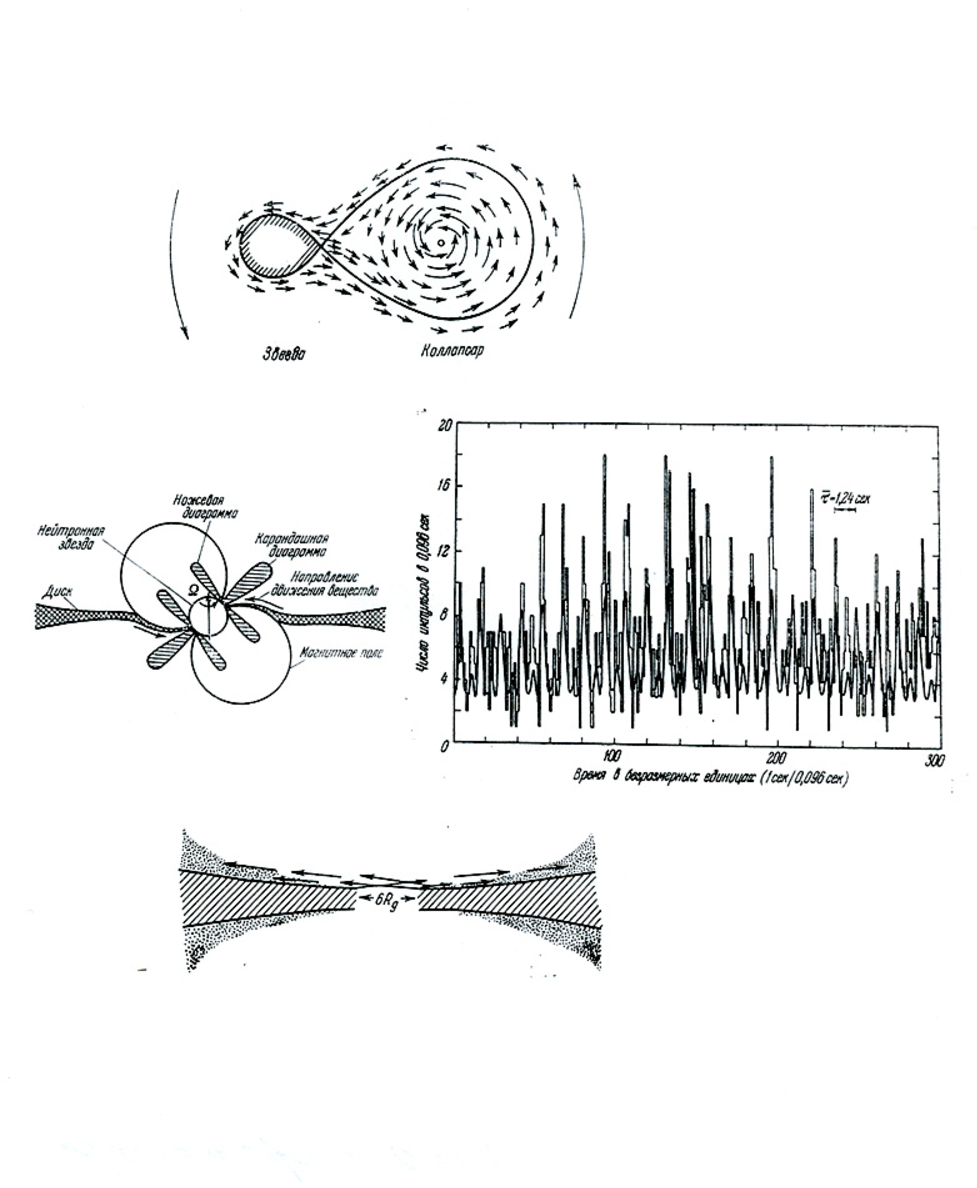}

\vspace*{-2cm}
\caption{Mathematical model of an X-ray binary system consisting
of an optical star and an accreting NS or BH. The optical light
curve of the Cyg X-1 X-ray binary is given below~\cite{LSCh73}.
Light variations are due to pear shape of the optical star and its
orbital motion (ellipticity effect). Orbital inclination, $i$, of
the binary system towards the picture plane is estimated using the
amplitude of light variations in the light curve.}
\label{Fig_2}
\end{figure}

A 3D gas-dynamic model of an X-ray binary system elaborated by A.A.Boyarchuk's team~\cite{BBK97} is displayed
in Fig.~\ref{Fig_2}. Powerful X-ray radiation is produced in the inner parts of the accretion disk close to
a BH.

\begin{figure} 
\centering\includegraphics[width=14cm]{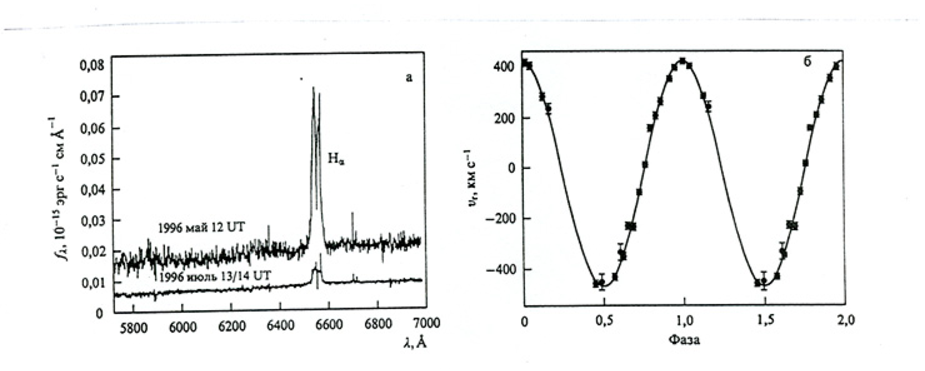} \caption{Optical spectrum of the
X-ray Nova -- the binary system with a BH, Nova Oph1977
(H1705-250), in a quiet state of radiation observed by Filippenko
et al.~\cite{FML97}. Numerous metallic absorption lines are seen
in the spectrum of K5V companion. Their Doppler shifts were used
to construct a radial velocity curve (shown right). There is also
$H_{\alpha}$ emission line of hydrogen displaying two-hump
structure formed in the accretion disk rotating around the BH.}
\label{Fig_3}
\end{figure}

Spectrum of optical radiation of X-ray Nova Oph1997 in a quiescent state~\cite{FML97} is shown in
Fig.~\ref{Fig_3}.

Measuring the doppler shifts of numerous metallic absorption lines
that correspond to a K-type star yields
the radial velocity curve representing projected orbital velocity of the star on the line of sight.
The radial velocity semi-amplitude of the optical star is about 400 km/s. Since radial velocities are measured
within 1-3 km/s error, the radial velocity curve is highly reliable. While interpreting it, however, one
should take into account that the optical star is not a material point, it has got considerable size and
is pear-shaped, temperature distribution over its surface being rather complicated. A mathematical technique
has been elaborated so far which allows for these effects to be taken into consideration~\cite{Ch03}.

\begin{figure} 
\centering\includegraphics[width=14cm]{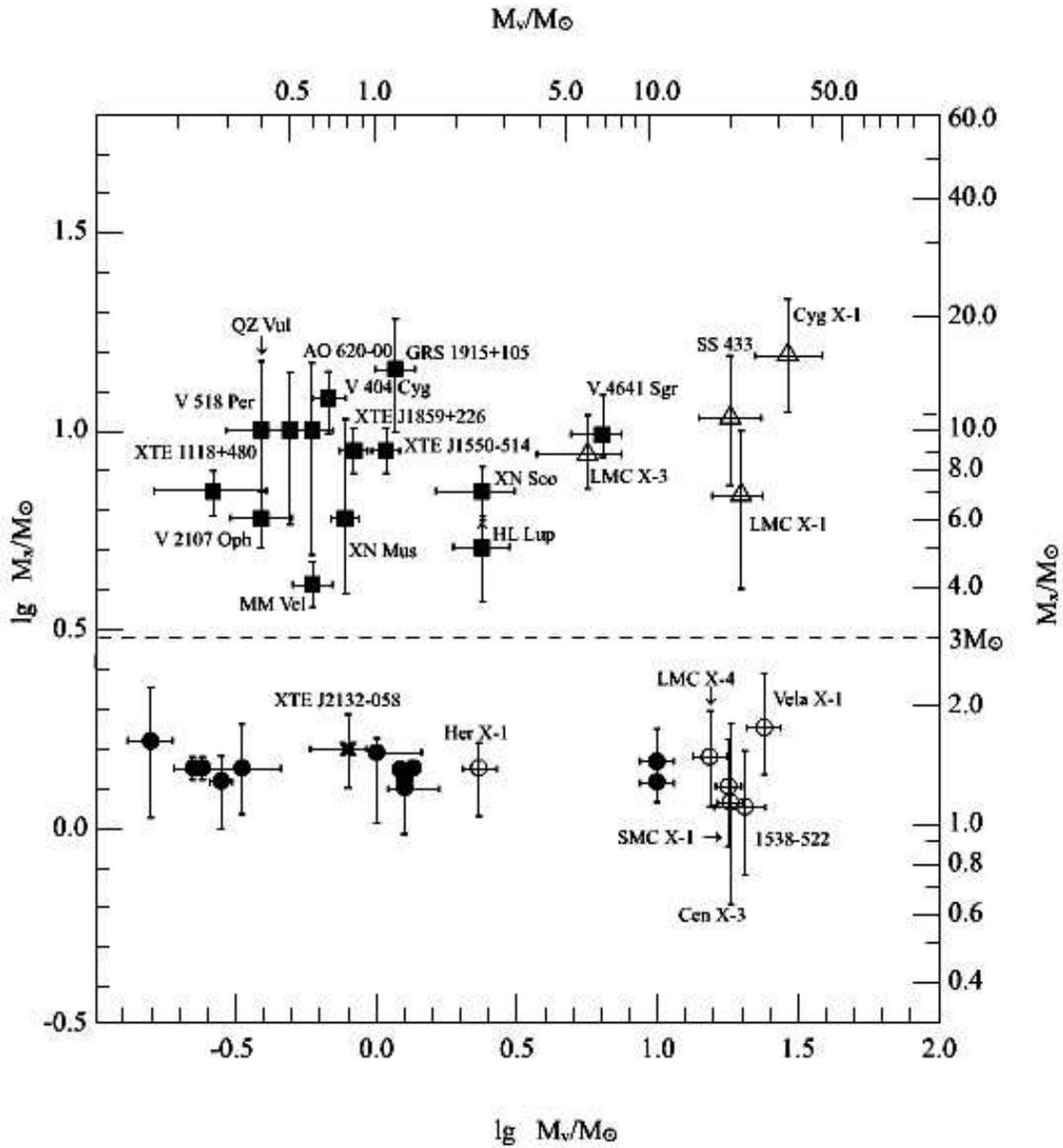} 

\vspace*{-3cm}
\caption{Masses of NS, $M_x$
(circles and crosses), and of BH (triangles and squares) versus
companion masses, $M_v$, in binary systems (in units of solar
masses $M_{\odot}$). Dark circles stand for radiopulsars, light     
ones for X-ray pulsars, crosses for the 1st type X-ray burster.
Dark squares correspond to BH in X-ray Novae, light triangles to
BH in quasistationary X-ray binary systems with hot massive
optical companions.} \label{Fig_4}
\end{figure}

Optical investigations of X-ray binary systems and measurements of
BH masses were carried out by American, Canadian and British
scientists (Charles, Cowley, Murdin, Hutchings, McClintock,
Remillard, Hynes, Martin, Casares, Orosz, Bailyn, Filippenko,
Shahbaz, Greiner, et al.) as well as Soviet and Russian researches
(Pavlenko, Lyuty, Sokolov, Fabrika, Goransky, Kurochkin, Shugarov,
et al.). The optical light curve of the X-ray binary system Cyg
X-1, the most plausible candidate for BH, first obtained by Lyuty,
Sunyaev and Cherepashchuk~\cite{LSCh73} is shown in
Fig.~\ref{Fig_1}. The amplitude of optical variability due to
optical star ellipticity effect gave inclination angle, i, between
the orbital plane and the picture plane and allowed one of the
earliest estimates of the BH mass to be obtained in Cyg X-1
system. A BH mass is calculated using the formula
$m_{BH}=f_{opt}(m)(1+q^{-1})^2 \sin ^{-3} i$ where $q=m_{BH}/m_{opt}$
is a ratio of the BH and optical star masses, $f_{opt}(m)$ is the
mass function of the optical star determined over its radial
velocity curve ($m_{BH}>f_{opt}(m)$). Parameters q and i are found
from additional information: optical light curve analyses,
rotational Doppler broadening of lines in the spectrum of the
optical star, distance to the binary system, duration of an X-ray
eclipse. The technique available is used to determine reliable BH
masses having spectral and photometric observations of X-ray
binary systems. Parameters for 18 X-ray binary systems with
measured BH masses are given in Table 1. Fig.~\ref{Fig_4} displays
masses of relativistic objects (NS and BH) versus companion masses
in the binaries.

Companions of X-ray pulsars,
1st type X-ray bursters, and BH in binary systems are O-M type optical stars. Companions of radiopulsars are
non-active NS, white dwarfs, and massive type ~B stars (we do not consider here radiopulsars with planets
as companions). Radiopulsar masses are determined at a high level of accuracy taking into account
re\-la\-ti\-vis\-tic effects in their orbital motions. Fig.~\ref{Fig_4} indicates that masses of relativistic objects
do not depend on masses of their companions: both NS and BH can belong to binary systems with massive and
low-mass companions. Orbital periods of X-ray binaries with BH are shown to lie in wide intervals between
0.17 days and 33.5 days. About a half of the systems have the mass function of the optical star more than
$3M_{\odot}$, i.e. the absolute upper limit for a BH mass predicted by General Relativity. In these cases
relativistic objects may be considered to be BH candidates, their masses being over $3M_{\odot}$. The
mathematical technique available allows one to find BH and NS masses, together with their errors, from
the mass function of the optical star (see Table 1). Masses of 19 NS are in the interval
$M_{NS}=(1-2)M_{\odot}$, a mean value of their mass being $(1.35\pm 0.15)M_{\odot}$. Masses of 18 BH have
been measured to lie in the interval $M_{BH}=(4-16)M_{\odot}$, a mean value of their mass being
$(8-10)M_{\odot}$.

The more numerous a number of relativistic objects with measured masses becomes (19 NS and 18 BH in the
Table 1), the stronger is the conviction that there is a systematic difference both in NS and BH masses and
in their observational manifestations according to Einstein's GR theory. For all objects with clearly
observed surface (radiopulsars, X-ray pulsars, or 1st type X-ray bursters) their masses (of NS) do not
exceed $3M_{\odot}$ what is in full accordance with GR. At the same time, among 18 massive ($M>3M_{\odot}$)
binary X-ray sources studied (BH candidates) there are neither radiopulsars nor X-ray pulsars, nor
1st type X-ray bursters. Hence, in full accordance with GR predictions, massive ($M>3M_{\odot}$) X-ray
sources, BH candidates, do not reveal any observable surface. A great number of relativistic objects
with measured masses (37) makes the conclusion rather reliable. It is an argument, though not a final
proof, that 18 BH candidates with measured masses are real BH in the sense of GR. Thus, presense or
absence of pulses or bursts is an observational manifestation crucial while defining whether or not an
accreting object is a NS or a BH. Moreover, there are finer spectral distinctions (in the range 1-10 KeV)
which indicate that NS have got surfaces observed whereas BH have not got~\cite{Ch03}.

Systems  GRS1915+105, SAX J1819.3-2525, GRO J1655-40, and 1E1740.7-2942 called microquasars display
relativistic collimated jets in X-ray bursts which have velocities $v\ge 0.92$ of the speed of light and
plasma cloud motions which have apparent velocities in excess of the speed of light (apparent superluminal
motions in the plane of the sky are due to the Special Theory of Relativity effects).

Recently interesting results have been obtained concerning
rotation of stellar-mass BH. If an accretion disk around the BH
rotates in the same direction as the BH itself, such a rotating
disk penetrates much closer to the BH than it would in the case of
a non-rotating BH. This is due to the fact that the radius of the
final stable orbit for a rotating BH is less than for a
non-rotating BH, $3r_g$.  Hence, luminosity and temperature of the
thermal component of X-rays  emitted by rotating accreting BH are
enhanced because of  more powerful release of energy in the
process of accretion. As a matter of fact, two transient X-ray
binary systems with BH, microquasars GRS1915+105 and GRO J1655-40,
display such enhanced characteristics and are likely to contain
fast rotating BH.

Radii of stellar-mass BH may be restricted while analyzing
observational results on fast variability of X-ray radiation. For
example, system Cyg X-1 displays fast irregular X-ray intensity
variability on a typical time scale $\Delta t$ up to $10^{-3}$ s.
A typical size of the region near a BH emitting X-rays is,
therefore, no more than $r=c \Delta t\cong 300$ {km/s} $\cong 10
r_g$. Observations of binary systems with BH have exhibited
wide-ranging quasiperiodic oscillations (QPOs). Detailed findings
on QPOs in X-ray binary systems with BH are given in the recent
review~\cite{McR03}. If high-frequency QPOs (typical frequencies
between 41 Hz and 450 Hz) are associated with orbital motions of
plasma condensations near a BH, the corresponding distancies are
not more than some gravitational radii. High-frequency QPOs may be
also connected with seismic oscillations of inner parts of the
accretion disk as the GR theory predicts or they may be caused by
relativistic dragging of the inertial reference frame near a fast
rotating BH.

\begin{figure} 
\centering\includegraphics[width=13cm,scale=0.7]{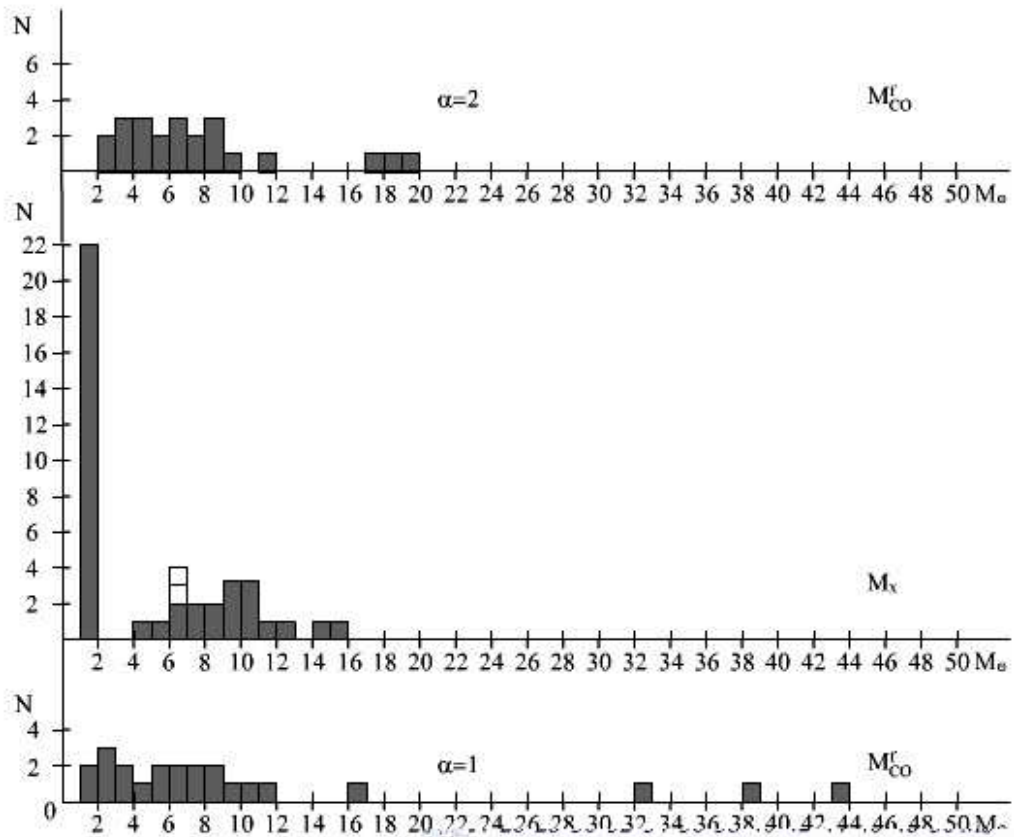} 

\vspace*{-3cm}
\caption{Observable
mass distribution of relativistic objects, $M_x$, in binary
systems is displayed in the middle. NS are concentrated in a
narrow mass interval, $M_x = (1-2)M_{\odot}$, while BH masses are
restricted by limits $4-16 M_{\odot}$. Masses of isolated BH
determined with the use of the microlensing effect are marked by
light squares. Mass distributions for the cores of WR stars,
$M^{f}_{CO}$, at the end of their evolution are shown at the
bottom and the top ($\alpha=1$ and $\alpha=2$ correspond to
different models of matter outflow from the star via stellar
wind). $M_x$ distribution is bimodal with the gap in the region
$M_x =(2-4)M_{\odot}$, whereas $M^{f}_{CO}$ distribution is             
continuous. Hence, the two mass distributions, those for
relativistic objects and for CO cores of massive stars at the end
of evolution, are of dramatically distinctive character.}
\label{Fig_5}
\end{figure}

Fig.~\ref{Fig_5} displays NS and BH masses versus mass
distributions, $M_{CO}^f$, of CO cores at the end of massive star
evolution (Wolf-Rayet stars).

Different models of stellar wind flowing from WR are designated
by parameters $\alpha=1$ and $\alpha=2$. Distribution of $M_{CO}^f$ in the interval $(1-12)M_{\odot}$ is seen
to be continuous while mass distribution of NS and BH resulting from the collapse of massive stellar CO-cores
is bimodal with two maxima and a dip in the interval $2-4M_{\odot}$. It seems like there is a deep reason
which would prevent the formation
of massive NS with $M>2M_{\odot}$ and low-mass BH with $M<4M_{\odot}$ in binary systems. Such an "avoidance
zone" for relativistic objects may be shown to be due to other reasons than observational selection. If the
conclusion is confirmed with more observational data, a more serious interpretation will be needed.

\section{Supermassive BH in Galactic Nuclei}

Most galaxies have got compact condensations of stars and gas in their centers which are called "nuclei".
Usually the cores are well visible in spiral galaxies and hardly discernible in irregular ones. Among all
galaxies a relatively small group may be singled out (~1 per cent) which involves galaxies with active
nuclei: Seyfert galaxies, radio galaxies, BL Lac galaxies, and quasars. The quasars are the most powerful
sources of stationary radiation in the Universe. Their total luminosity reaches $10^{47}$ erg/s, which is
three orders of magnitude more than that of a host galaxy. Intense non-stationary processes occuring in
active nuclei result in variability of optical radiation on time scales of days to many years. Spectra of
these active nuclei exhibit strong and often wide emission lines of hydrogen, helium and other elements.
Many nuclei of active galaxies are observed to have strongly collimated jets of matter moving with
relativistic velocities. A galactic nucleus is currently considered to be a supermassive BH with accretion
of stellar matter and gas~\cite{R82}.

To determine BH masses in galactic nuclei a hypothesis is used according to which the gravitational field of
a central object controls the motion of gas and stars near the nucleus~\cite{D84}. As was mentioned above,
Newton's law can be used since $r>r_g$. In this case the velocity $v$ of a star or a gas cloud depends on
the distance $r$ to the center of a galaxy as $v^2\sim r^{-1}$. Hence the BH mass in the nucleus can be
estimated as $M_{BH}=\eta v^2$ r/G, where $\eta=1-3$ depending on a kinematic model of body motion
around the galactic center (for circular motion, $\eta=1$).

It is possible in many cases to see the moving gas directly, for our Galaxy even individual stars, near
the galactic center owing to modern observational facilities (the Hubble Space Telescope, very large
groundbased
telescopes provided with techniques for compensation of atmospheric distortions, intercontinental
interferometers, etc.). Therefore the BH masses are determined unequivocally using the formula given.

If a disk of gas and dust surrounding the galactic center cannot be seen and its rotation cannot be
investigated another method is applied based on statistical researches of stellar kinematics in the
central parts of the galaxy which is defined by the BH gravitational influence.

BH masses in active galactic nuclei with observed strong and wide emission lines can be determined using the
formula given. Velocities, $v$, of gas clouds near the nucleus which are responsible for a wide component
of emission line profiles can be estimated with the help of the Doppler semi-amplitude of this wide line
component. The distance, $r$, of gaseous clouds to the nucleus can be estimated by two ways: by means of a
photoionization model of the nucleus~\cite{D84, Sh65} or by time delay, $\Delta t$, which reveals in
the fast variability of a wide emission line component with regard to the variability of the continuous
spectrum, $r\approx c \cdot\Delta t$ (so called reverberation mapping). The time delay effect was discovered by
Cherepashchuk and Lyutyi in 1973~\cite{ChL73}. It was mentioned there that the time delay of the line
variability
with regard to the continuous spectrum is equal to the time needed for ionizing radiation to cover the
distance from the galactic center to the gaseous clouds emitting in the line. Thus an independent estimate of
the distance
can be obtained  and the mass of an active galaxy nucleus can be reliably determined.

The first method to estimate BH masses in active galactic nuclei
suggested that the bolometric luminosity of the nucleus should be
close to the Eddington limit~\cite{ZN64}. Such estimates give the
following values for quasar nuclei masses: $M>10^8 M_{\odot}$.

So far there are some dozens of BH masses in active galactic
nuclei estimated by means of the time delay effect according to
which emission line variability lags behind continuum~\cite{Ch03}.

``Normal galaxies'' have got nuclei which are characterized by
rather faint optical activity as compared to their stellar
constituent. In such galaxies stars and gas moving near the
nucleus can be observed directly what permits the most exact and
model-independent mass estimates of supermassive BH to be
obtained. Recently disks of gas and dust spanning some dozens or
hundreds of parsecs around the nuclei of many galaxies and
rotating according to Newton's law were discovered aboard the
Hubble space telescope with high angular resolution
(see~\cite{M99} and references therein). To check the validity of
the Keplerian rotation law for a disk ($v \sim r^{-1/2}$) and to obtain
the inclination angle $i$ between the disk axis and the line of
sight Doppler shifts are investigated in emission lines using the
projection of the near-nucleus region of the disk onto the picture
plane. Then the mass in the volume with radius $r$ is estimated
unequivocally. Since the near-nucleus region in the galaxy may be
observed directly the mass-to-luminosity ratio, $M(r)/L(r)$, can
be estimated and compared with the corresponding value for
external parts of the galaxy, $M/L\cong 1\div 10$, where $M$ and
$L$ are the solar mass and luminosity respectively. The first
galaxy to have been used for determining the mass of the central
BH using the near-nucleus disk of gas and dust with a luminous and
stretched jet was M87~\cite{FHT94}. The mass of the central BH is
$(3.2\pm 0.9)\cdot 10^9 M_{\odot}$, the mass-to-luminosity ratio
is $M/L>110$. If the central mass was due to a dense cluster of
ordinary stars rather than a BH, the nucleus of M87 would be
dozens of times brighter than what is actually observed. An
average density of dark matter in the nucleus of M87 is estimated
to be $~10^7 M_{\odot}/\mbox{pc}^3$ whereas star density in
external parts of the galaxy is $~0.5 M_{\odot}/\mbox{pc}^3$ and
in the most dense stellar clusters $~10^5 M_{\odot}/\mbox{pc}^3$.
All these findings allow one to believe with a good reason that
there is a supermassive BH in the nucleus of M87 (three billion
solar masses) which undergoes accretion of matter causing many
aspects of M87 activity including formation of a relativistic jet.
A number of mass estimates obtained so far for supermassive BH by
means of researching gas and star kinematics near galactic nuclei
reaches many dozens (see, for example, review~\cite{Ch03}). Table
2 gives some results on determining BH masses in galactic nuclei.

Outstanding results for estimating BH masses in galactic nuclei
have been obtained recently while researching compact maser
sources in near-nucleus molecular disks by means of
intercontinental radioastronomy (see review by Moran et
al.~\cite{MGH99} and references therein). Observations of NGC4258
galactic nucleus revealed 17 compact maser sources emitting very
narrow $H_2 O$ lines and being located in a disk-like envelope
with a radius of $~10^{17}$ cm which is seen nearly edge-on.
Velocities of maser sources are distributed according to the
Keplerian law. The mass of a central BH is $3.9\cdot 10^7
M_{\odot}$. This method has been used so far to measure masses of
about a ten BHs in nuclei of galaxies (see reviews~\cite{MGH99,
Ch03} and references therein).

\section{A Supermassive Black Hole in the Nucleus of Our Galaxy}

\begin{figure}[th] 
\centering\includegraphics[width=13cm]{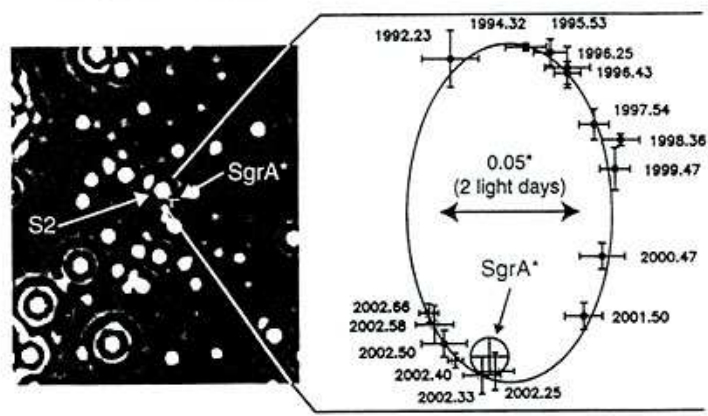} %
\vspace*{-1cm}
\caption{The orbit of S2 star moving around the supermassive BH in
the center of our Galaxy constructed by R.Schoedel et
al.~\cite{SOG02}. To the left: the sky region in the field of
SgrA* source, the center of the Galaxy, which is seen to contain a
cluster of stars.} \label{Fig_6}
\end{figure}

The most convincing evidence for a supermassive BH have been
obtained recently while researching motions of individual stars in
the closest surroundings of SgrA* source, the center of our
Galaxy. Beginning from the 90s of the past century the motions of
individual stars have been investigated in the picture plane near
the center of our Galaxy~\cite{EG96}. Observations are carried out
in the IR range using special techniques for compensation of
atmospheric distortions of the image (the Galactic Center is
hidden from optical view by thick layers of gas and dust). The
stars near the Galactic Center are found to shift considerably,
their velocities being the higher the closer to the Center.
Recently R.Schoedel~\cite{SOG02} et al. constructed an orbit of
one of the closest stars to the Galactic Center (SO2) (see
Fig.~\ref{Fig_6}).

\vspace*{-1cm}

The star SO2 has 15.2-year orbit with an eccentricity of 0.87 and a semi-major axis of $4.62\cdot 10^{-3}$ pc
($~20 000 r_g$). Kepler's third law yields, for a BH mass, $(3.7\pm 1)\cdot 10^6 M_{\odot}$. The
dark gravitating matter density in the field measured reaches $10^{17} M_{\odot}/\mbox{pc}^3$ while a typical
dynamical break-up time of a supposed cluster of individual dark bodies in the galactic nucleus (due to
collective collisions) is estimated to be $~10^5$ years, the age of the Galaxy being $10^{10}$ years. This
argues strongly that the massive compact object in the center of the Galaxy forms a whole dark body rather
than a cluster of individual low-mass objects. Moreover, orbits of eight individual stars near the galactic
center were measured lately: SO-16, SO-19, SO-20, SO-1, SO-2, SO-3, SO-4, SO-5. The BH mass in the Galactic
nucleus is estimated to be $(4\pm 0.3)\cdot 10^6 M_{\odot}$, it is situated at the dynamic center of the Galaxy
within $\pm 10^{-3}$ arcsec. The BH proper motion is $~(0.8\pm 0.7)\cdot 10^{-3} \mbox{s}\cdot\mbox{year}^{-3}$ what is actually zero within the limits of error. These findings argue strongly in favour of
Gurevich's idea~\cite{IZG03} that supermassive BH form in galactic nuclei due to accretion of baryonic matter
which falls into potential wells in the center of galactic halos of dark matter. The SO-16 star
approaches the BH as close as 90 AU ($1700 r_g$) while moving in the orbit around.

\section{Observational Restrictions upon Radii of Black Holes in Galactic Nuclei}

According to X-ray image data with  resolution $0.''5$ obtained by the Chandra observatory the center of
the Galaxy is shown to emit variable X-rays. On the time scale of a year X-ray luminosity changes between
$2\cdot 10^{33}$ and $10^{35}$ erg/s, the galactic nucleus displaying rapid variability (as high as 5 times
on the scales $t_{min}\le 10$ min)~\cite{BBB01}. Hence the size of the region emitting in X-rays,
$r\le ct_{min}$, is $20 r_g$. The center of the Galaxy shows  large variations in IR flux density,
a factor of 2 over 40 min, as was discovered with the W. M. Keck II 10-meter telescope~\cite{GWM03}.
This variability implies that the size of the IR emitting field does not exceed $80 r_g$. IR luminosity
of the galactic center is $~10^{34}$ erg/s at 3.8 $\mu$m. The variable IR source is coincident with the
center of the Galaxy to within $6\cdot 10^{-3}$ arcsec and does not move, its velocity being at least
$v<300$ km/s, whereas  the stars near the galactic center have got constant luminosity and move around
the center with velocities of thousands of km/s.

Therefore, observations indicate that there is a massive compact object with a mass of $4\cdot 10^6 M_{\odot}$
and a radius less than 20 gravitational radii in the center of our Galaxy, the parameters arguing in favour of
the object being a supermassive BH.

Direct measuring a supermassive BH radius in the center of the Galaxy (as well as in the centers of nearby
galaxies) will be possible after launching space interferometers: an X-ray interferometer with
$10^{-7}$ arcsec resolution~\cite{W00} and the RadioAstron interferometer with $10^{-6}$ arcsec resolution
in the radio range. Angular sizes of the supermassive BH in the centers of our Galaxy and the Andromeda
Galaxy are $~7\cdot 10^{-6}$ and $~3\cdot 10^{-6}$ arcsec, respectively. Launching these interferometers
will allow not only supermassive BH radii to be measured but also physical phenomena to be observed connected
with plasma moving near the event horizon. Such experiments are likely to provide sufficient criteria to
select BH and to prove, once and for all, their existence in the Universe. Using contemporary intercontinental
interferometry techniques made it possible to research, in the millimeter range, the formation of jet in the
inner parts of the M87 galaxy as well as to confine directly a value of the supermassive BH radius within
$r<30-100$ gravitational radii~\cite{JBL99}.

\begin{figure}[th] 
\centering\includegraphics[width=11cm]{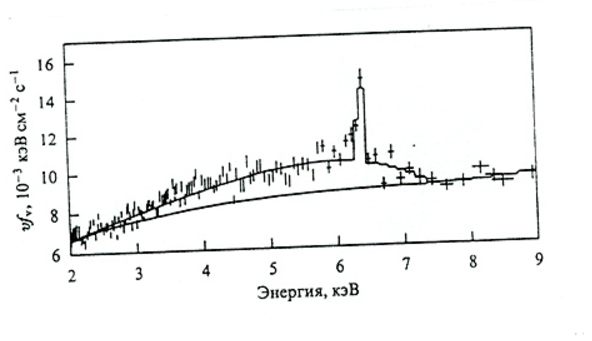} \caption{The profile of Fe X-ray
$K_{\alpha}$ line in the spectrum of the Seyfert galaxy
MCG-6-30-15 obtained by the X-ray observatory XMM in the period of
low X-ray nucleus luminosity (Wilms J. et
al.~\cite{WRB01}). Along with a narrow line with the energy close
to the laboratory standard, $\sim 6.4$ keV, a very large profile
component is observed which is greatly displaced towards the low
spectral energies. This indicates that the region of line
formation is within three gravitational radii from the central
supermassive BH.} \label{Fig_7}
\end{figure}

In addition, iron $K_{\alpha}$ emission line profile in X-rays at ~6.4 keV in spectra of active galactic
nuclei observed aboard X-ray observatories ASCA, CHANDRA, and XMM with high spectral resolution~\cite{WRB01}
also restricts strongly the values of BH radii. Relativistic  effects near the event horizon of the central BH
result in the redshift of a spectral line, its specific asymmetric profile and a huge width (up to ~100,000
km/s) thus providing bounds on the supermassive BH radius in the center of a galaxy. For instance, in case of
MCG-6-30-15 galactic nucleus, analysis of the wide spectral component of Fe XXV X-ray line profile indicates
that the inner edge of the accretion disk is located less than $3r_g$  from the central supermassive BH which
seems to rotate~\cite{WRB01} (see Fig.~\ref{Fig_7}).

\section{Demography of Supermassive Black Holes}

A number of supermassive BH with their masses measured approaches now 200. The radii estimated are available
for many of them: $r<(10-100)r_g$. Hence a new field in astrophysics is intensively developing now,
the de\-mo\-gra\-phy of BH. The basic results achieved in this field are briefly stated below.

\begin{enumerate}

\item Supermassive BH in galactic nuclei have masses correlated with those of galactic bulges, spherical
condensations of old low-mass stars near the nucleus with the large dispersion of velocities~\cite{McLD02}:\\
$M_{BH}=0.0012 M^{0.95\pm 0.05}_{bulge}$.
\item There is the correlation between supermassive BH masses and velocity dispersion of stars, $\sigma$,
belonging to the bulge~\cite{TGB02}: $M_{BH} \sim \sigma ^4 _{bulge}$.
\item Masses of supermassive BH are correlated with linear rotation velocities of galaxies in the range
of a costant value of the velocity.  Since a linear rotation velocity of a galaxy far away from its center
is mainly due to the
gravitational pull of the galactic halo consisting of dark matter, the mass of the central
supermassive BH should be correlated with the mass of the galactic halo~\cite{BBH03}: $M_{BH} \sim
M^{1.27}_{halo}$. The result obtained is an important evidence in favour of Gurevich's model~\cite{IZG03}.

\end{enumerate}

\section{Conclusion}

The actual state of the problem connected with searching for both stellar-mass BH and  supermassive BH
in the galactic nuclei has been described. We did not touch upon the problem concerning searches for primary BH
because of the lack of observational data and ambiguity revealed in their interpretation. Primary BH may
as well exist among isolated  stellar-mass BH. It is worthwhile mentioning that, using gravitational
microlensing effect \cite{MSW02}, the masses of two isolated BH have been measured so far: $m_{BH}\cong
6 M_{\odot}$ (the brightness of a distant background star being strenghened because of the influence of
a foreground object playing a part of a "gravitational lense", the duration of its strengthened brightness
is proportional to a square root of the gravitational lense mass). The problem for searching BH of
intermediate mass ($m_{BH}=10^2 \div 10^4 M_{\odot}$) was not considered either because there were not
convincing findings in this field. The review on intermediate mass BH located both in galaxies and in
stellar clusters is available in~\cite{M03}.

It is important that the problem with searching for BH is actually sup\-port\-ed with a firm observational basis
and a number of BH being discovered is gradually increasing (around 200 at the present moment). It should be
espe\-ci\-al\-ly emphasized that all necessary constraints imposed upon BH ma\-ni\-fes\-ta\-ti\-ons by
Einstein's General Relativity
are fulfilled as it follows from observational data. This strengthens considerably our confidence in actual
existence of BH in the Universe.

The main task to be solved in the decade to come is to find sufficient criteria for BH  candidates being
real BH. The following experiments may be expected to help in solving this problem:

\begin{enumerate}
\item Using space interferometers with $~10^{-6} - 10^{-7}$ arcsec angular resolution and direct observations of
matter moving close to event horizons of supermassive BH in nuclei of our and nearest galaxies.
\item Searches and investigations of gravitational wave bursts from BH coales\-cen\-ce in binary systems with
the use of laser gravitational wave in\-ter\-fe\-ro\-met\-ry (LIGO, VIRGO, LISA, etc.).
\item Detection and researches of radiopulsar motion in binary systems with BH (among ~1000 pulsars
one pulsar is anticipated to be paired with a BH, about 1500 pulsars being known so far).
\item Detailed investigations of spectra, intensity, polarization and variability of X-ray and gamma-radiation
from accreting BH with the use of new generation orbital observatories.
\item Observations and interpretations of gravitational microlensing effects in galactic nuclei caused by
closer galaxies (gravitational lenses).
\item Routine  storage of evidence concerning masses of black holes and neutron stars (NS) as well as
statistical comparison between observational ma\-ni\-fes\-ta\-ti\-ons  of BH and NS.
\end{enumerate}

The work is supported by the Russian Fund for Fundamental Research (grant 02-02-17524).


\renewcommand{\refname}{References}

\newpage

\begin{center}
{\bf TABLE 1. Parameters of Binary Systems with Black Holes}
\end{center}

{\small
\begin{tabular}{|l|l|c|c|c|c|c|l|}
\hline
System &  Opt. Star & $P_{orb}$ & $f_{opt}(M)$ & $M_{BH}$    & $M_{opt}$   & $V_{pec}$ & Note\\
       &    Spectrum  & (days)    & $(M_{\odot})$  & $(M_{\odot})$ & $(M_{\odot})$ & (km/s)    &      \\
\hline
Cyg X-1 & O 9.7 Iab    & 5.6     & \,\,\,0.24 & \,\,16   & \,\,33  & \,\,\, 49& stat.\\
V 1357 Cyg & & & $\pm 0.01$ & $\pm 5$ & $\pm 9$  & $\pm 14$ & \\
\hline
LMC X-3 & B3 Ve    & 1.7     & \,\,\,2.3 & \,\,\,9   & \quad6  & - & stat.\\
 & & & $\pm 0.3$ & $\pm 2$ & $\pm 2$  & & \\
\hline
LMC X-1 & O(7-9) III    & 4.2     & \quad0.14 & \,\,\,7   & \,22  & - & stat.\\
& & & $\pm 0.05$ & $\pm 3$ & $\pm 4$  & & \\
\hline
SS 433 & $\sim$ A7 Ib    & 13.1     & $\sim 1.3$ & \,\,11   & 19  & - & stat.\\
& & & & $\pm 5$ & $\pm 7$  & & \\
\hline
A0 620-00 & K5 V    & 0.3     & \quad2.91 & \,\,10   & \quad0.6  & -15 & trans.\\
(V 616 Mon) & & & $\pm 0.08$ & $\pm 5$ & $\pm 0.1$ & $\pm 5$ & \\
\hline
GS 2023+338 & K0 IV    & 6.5     & \,\,\,6.08 & \,\,12   & \quad0.7  & \,\,\,8.5 & trans.\\
(V 444 Cyg) & & & $\pm 0.06$ & $\pm 2$ & $\pm 0.1$  & $\pm 2.2$ & \\
\hline
GRS 1124-68 & K2 V    & 0.4     & \,\,\,3.01 & 6   & \quad0.8  & 26 & trans.\\
(GU Mus) & & & $\pm 0.15$ & (+5,-2) & $\pm 0.1$ & $\pm 5$ & \\
\hline
GS 2000+25 & K5 V    & 0.3     & \,\,\,4.97 & \,10   & \quad0.5  & - & trans.\\
(QZ Vul) & & & $\pm 0.10$ & $\pm 4$ & $\pm 0.1$ & & \\
\hline
GRO J0422+32 & M2 V    & 0.2     & \quad1.13 & \,10   & \quad0.4  & - & trans.\\
(V 518 Per) & & & $\pm 0.09$ & $\pm 5$ & $\pm 0.1$ & & \\
\hline
GRO J1655-40 & F5 IV    & 2.6     & \quad2.73 & \quad6.3   & \quad2.4  & -114 & trans.\\
(XN Sco 1994) & & & $\pm 0.09$& $\pm 0.5$  & $\pm 0.4$ & $\pm 19$ & \\
\hline
H 1705-250 & K5 V    & 0.5     & \quad4.86 & $6\pm 1$   & \quad0.4  & \quad38 & trans.\\
(V 2107 Oph) & & & $\pm 0.13$ & & $\pm 0.1$ & $\pm 20$ & \\
\hline
4U 1543-47 & A2 V    & 1.1     & \quad0.22 & \,\,4.0-  & $\sim 2.5$  & - & trans.\\
(HL Lup) & & & $\pm 0.02$ & 6.7 & & & \\
\hline
GRS 1009-45 & (K6-M0) V    & 0.3     & \quad3.17 & \,\,\,3.6-   & \,\,0.5-  & - & trans.\\
(MM Vel) & & & $\pm 0.12$ & 4.7 & 0.7 & & \\
\hline
SAX J1819.3-2525 & B9 III    & 2.8     & \quad2.74 & 9.61   & 6.53  & - & trans.\\
(V 4641 Sgr) & & & $\pm 0.12$ & (+2.08- & (+1.6- & & \\
& & & & \quad\,0.88) & 1.03) & & \\
\hline
XTE 1118+480 & (K7-M0)V    & 0.17     & \,6.0 &    & \quad0.09-  & 126 & trans.\\
& & & -7.7 & & 0.5 & & \\
\hline
GRS 1915+105& (K-M)III    & 33.5     & \quad9.5 & $14\pm 4$   & \quad1.2  & - & trans.\\
& & & $\pm 3.0$ & & $\pm 0.2$ & & \\
\hline
XTE J1550-564 & $\sim K3$    & 1.54     & \quad6.86 &  \quad8.36-  & $\sim 1$  & - & trans.\\
& & & $\pm 0.71$ & 10.76 & & & \\
\hline
XTE J1859+226 & $\sim K7$    & 0.38     & \quad7.4 &  \quad7.6-  & $\sim 0.7$  & - & trans.\\
& & & $\pm 1.1$ & 12.0 & & & \\
\hline
\end{tabular}

\newpage

\textbf{Note}: $P_{orb}$ stands here for an orbital period,
$f_{opt}(M)=\frac{M^3_{BH}\sin^3 i}{(M_{BH}+M_{opt})^2}$ is a mass
function of an optical star, $M_{BH}$, $M_{opt}$ are masses of
black holes and optical companions, respectively, $V_{pec}$ is a
peculiar velocity of the center of gravity of a binary system. }

\enddocument
\newpage
\begin{center}
{\bf TABLE 2. Masses of Supermassive Black Holes in Galactic Nuclei Estimated from Kinematics of Gas and Stars}
\end{center}

{\small
\newcommand{\PreserveBackslash}[1]{\let\temp=\\#1\let\\=\temp}
\let\PBS=\PreserveBackslash
\begin{longtable}
{|p{80pt}|p{20pt}|p{50pt}|p{30pt}|p{30pt}|p{80pt}|p{50pt}|} a & a
& a & a & a & a & a  \kill \hline ч┴╠┴╦╘╔╦┴& Ї╔╨& Є┴╙\-╙╘╧╤\-╬╔┼
\par э╨╙& M$_\textrm{B}$ \par (bulge)& $\sigma_{1}$ \par ╦═/╙&
э$_\textrm{▐.─.}$(max,min) \par (э$_\odot$)&
э┼╘╧─ \par ╧╨╥┼\-─┼╠┼\-╬╔╤ \\
\hline Milky Way& SBbc& 0.008& -17.65& 103& 1.8$ \cdot
$10$^{6}$(1.5,2.2)&
s,p \\
\hline NGC221 =M32& E2& 0.81& -15.83& 75& 2.5$ \cdot
$10$^{6}$(2.0,3.0)& s,3I \\
\hline NGC224 =M31& Sb& 0.76& -19.00& 160& 4.5$ \cdot
$10$^{7}$(2.0,8.5)& s \\
\hline NGC821& E4& 24.1& -20.41& 209& 3.7$ \cdot
$10$^{7}$(2.9,6.1)& s,3I \\
\hline NGC1023& SB0& 11.4& -18.40& 205& 4.4$ \cdot
$10$^{7}$(3.9,4.9)& s,3I \\
\hline NGC1068& Sb& 15.0& -18.82& 151& 1.5$\cdot
$10$^{7}$(1.0,3.0)& m \\
\hline NGC2778& E2& 22.9& -18.59& 175& 1.4$ \cdot
$10$^{7}$(0.5,2.2)& s,3I \\
\hline NGC2787& SB0& 7.5& -17.28& 140& 4.1$ \cdot
$10$^{7}$(3.6,4.5)& g \\
\hline NGC3115& S0& 9.7& -20.21& 230& 1.0$ \cdot
$10$^{9}$(0.4,2.0)& s \\
\hline NGC3245& S0& 20.9& -19.65& 205& 2.1$ \cdot
$10$^{8}$(1.6,2.6)& g \\
\hline NGC3377& E5& 11.2& -19.05& 145& 1.0$ \cdot
$10$^{8}$(0.9,1.9)& s,3I \\
\hline NGC3379& E1& 10.6& -19.94& 206& 1.0$ \cdot
$10$^{8}$(0.5,1.6)& s,3I \\
\hline NGC3384& S0& 11.6& -18.99& 143& 1.6$ \cdot
$10$^{7}$(1.4,1.7)& s,3I \\
\hline NGC3608& E2& 22.9& -19.86& 182& 1.9$ \cdot
$10$^{8}$(1.3,2.9)& s,3I \\
\hline NGC4258& Sbc& 7.2& -17.19& 130& 3.9$ \cdot
$10$^{7}$(3.8,4.0)& m,a \\
\hline NGC4261& E2& 31.6& -21.09& 315& 5.2$ \cdot
$10$^{8}$(4.1,6.2)& g \\
\hline NGC4291& E2& 26.2& -19.63& 242& 3.1$ \cdot
$10$^{8}$(0.8,3.9)& s,3I \\
\hline NGC4342& S0& 15.3& -17.04& 225& 3.0$ \cdot
$10$^{8}$(2.0,4.7)& s,3I \\
\hline NGC4459& S0& 16.1& -19.15& 186& 7.0$ \cdot
$10$^{7}$(5.7,8.3)& g \\
\hline NGC4473& E5& 15.7& -19.89& 190& 1.1$ \cdot
$10$^{8}$(0.31,1.5)& s,3I \\
\hline NGC4486 =M87& E0& 16.1& -21.53& 375& 3.0$ \cdot
$10$^{9}$(2.0,4.0)& g \\
\hline NGC4564& E3& 15.0& -18.92& 162& 5.6$ \cdot
$10$^{7}$(4.8,5.9)& s,3I \\
\hline NGC4596& SB0& 16.8& -19.48& 152& 7.8$ \cdot
$10$^{7}$(4.5,12)& g \\
\hline NGC4649& E1& 16.8& -21.30& 385& 2.0$ \cdot
$10$^{9}$(1.4,2.4)& s,3I \\
\hline NGC4697& E4& 11.7& -20.24& 177& 1.7$ \cdot
$10$^{8}$(1.6,1.9)& s,3I \\
\hline NGC4742& E4& 15.5& -18.94& 90& 1.4$ \cdot
$10$^{7}$(0.9,1.8)& s,3I \\
\hline NGC5845& E3& 25.9& -18.72& 234& 2.4$ \cdot
$10$^{8}$(1.0,2.8)& s,3I \\
\hline NGC6251& E2& 93.0& -21.81& 290& 5.3$ \cdot
$10$^{8}$(3.5,7.0)& g \\
\hline NGC7052& E4& 58.7& -21.31& 266& 3.3$ \cdot
$10$^{8}$(2.0,5.6)& g \\
\hline NGC7457& S0& 13.2& -17.69& 67& 3.5$ \cdot
$10$^{6}$(2.1,4.6)& s,3I \\
\hline IC1459& E3& 29.2& -21.39& 340& 2.5$ \cdot
$10$^{9}$(2.1,3.0)& s,3I \\
\hline
\end{longtable}


\textbf{Note:} $M_B(bulge)$ is here a stellar B magnitude of the
galactic bulge, $\sigma_1$, the velocity dispersion of bulge
stars, $M_{BH}$, the mass of the central BH in terms of
$M_{\odot}$ (with maximum and minimum BH mass values in brackets
corresponding to RMS deviations of mass determination).
Designations in the last column stand for different ways of BH
mass determination: s - stellar radial velocities, p - stellar
proper motions, m - radial velocities of gas clouds estimated over
maser emission lines, g - a rotating gas disk observed in emission
lines, 3I - axially symmetric dynamical model involving three
integrals of the motion.}




\bigskip


\end{document}